\documentclass[%
 reprint
]{revtex4-2}

\usepackage[T1]{fontenc}
\usepackage[latin9]{inputenc}
\usepackage{babel}
\usepackage{nccmath}
\usepackage{orcidlink}
\usepackage{placeins}

\usepackage{graphicx,amsmath,amssymb,color}
\usepackage[normalem]{ulem}
\usepackage{amsfonts}
\usepackage[toc,page]{appendix}
\usepackage{url} 
\usepackage{latexsym}
\usepackage{amsfonts}
\usepackage{algpseudocode}
\usepackage{amsthm}
\usepackage{mathrsfs}
\usepackage{natbib}
\usepackage{color,verbatim}
\usepackage{psfrag}
\usepackage{hyperref}
\usepackage{cleveref}

\newcommand{\resp}[1]{{\color {black} #1 }}
\newcommand{\resptwo}[1]{{ #1 }}

\renewcommand{\cref}{Fig.~\ref}
\def\beq{\begin{equation}}
\def\eeq{\end{equation}}
\def\bali{{\begin{align}}}
\def\eali{{\end{align}}}
\def\ie{{\it i.e.~}}
\def\eg{{\it e.g.~}}
\def\ODM{\Omega_{\rm DM}}

\preprint{IPPP/23/XX} 

\begin{document}

\title{Symbolic Regression for Beyond the Standard Model Physics}

\author{Shehu AbdusSalam$^{(1)}$\orcidlink{0000-0001-8848-3462}}
\email{abdussalam@sbu.ac.ir}

\author{
Steven Abel$^{(2)}$\orcidlink{0000-0003-1213-907X}}
\email{s.a.abel@durham.ac.uk}

\author{Miguel Crispim Rom\~ao$^{(2)}$\orcidlink{0000-0003-4539-6283}}
\email{miguel.romao@durham.ac.uk}

\address{\vspace{0.5cm} 
\hspace{-0.2cm}$^{(1)}$
Department of Physics, Shahid Beheshti University, Tehran, Iran\\
$^{(2)}$Institute for Particle Physics Phenomenology, Department of Physics, Durham University, Durham DH1 3LE, U.K.}

\begin{abstract}
We propose symbolic regression as a powerful tool for the numerical studies of proposed models of physics Beyond the Standard Model. In this paper we demonstrate the efficacy of the method on a benchmark model, namely the Constrained Minimal Supersymmetric Standard Model, which has a four-dimensional parameter space. We provide a set of analytical expressions that reproduce three low-energy observables of interest in terms of the parameters of the theory: the Higgs mass, the contribution to the anomalous magnetic moment of the muon, and the cold dark matter relic density. To demonstrate the power of the approach, we employ the symbolic expressions in a global fits analysis to derive the posterior probability densities of the parameters, which are obtained two orders of magnitude more rapidly than is possible using conventional methods. 
\end{abstract}

\maketitle

The chief test of any proposal for Beyond the Standard Model (BSM) physics is to confront it with experimental data. The standard approach is well trodden: first one chooses a reasonable parameter space motivated by a combination of physical argumentation and expediency; for each point in the parameter space the physical low-energy spectrum is determined and possibly an initial cut made for phenomenological viability (for example in supersymmetry (SUSY) the mass and charge of the lightest SUSY partner); for each remaining viable point the cross-sections are calculated and the relevant observables determined such as dark matter relic density, anomalous magnetic moment of the muon, and so forth; finally with this information to hand each point can be evaluated and assessed. One can then attempt to scan the entire parameter space this way, or alternatively use a Markov Chain Monte Carlo or nested sampling algorithm~\cite{Skilling:2004pqw, Skilling:2006gxv, Ashton:2022grj}, such as that in \resptwo{MultiNest~\cite{Feroz:2007kg} and Dynesty~\cite{2020MNRAS.493.3132S}}, to arrive at posterior probability densities for the parameters.

This approach has the appeal of being directly connected to the underlying physics, but it suffers from a severe bottleneck, namely the computation of the observables. Indeed each one of them is the result of a painstaking physical analysis which may need to encompass many subtle effects (for example, three loop running from the Grand Unified Theory (GUT) scale, co-annihilation for dark matter relic density and so forth). Typically, this leads to a chain of computation to get from the input parameters to the low-energy observables. Thus, although closed-form expressions for the observables in terms of the input parameters are in principle ``knowable'' (at least in perturbation theory), they would be exceedingly complex and could not be usefully expressed analytically except possibly in the case of a restricted set of observables in extreme limits of parameter space. 

To avoid this bottleneck, it is natural to turn to machine learning to bypass the computation chain or more efficiently sample points of interest (for recent examples and applications, see Refs.~\cite{Caron:2016hib,Hammad:2022wpq,deSouza:2022uhk,Romao:2024gjx,Diaz:2024yfu}). However, the negative aspect of machine learning is that it is generally neither interpretable nor explainable. 

This lack of interpretability (by which we mean an inability to be able to understand the dependence of the observables on the input parameters) is frustrating because certain correlations between input parameters and observables can be motivated by physical arguments. For example, it is clear that SUSY contributions to $(g-2)_\mu$, the anomalous magnetic moment of the muon, generally decrease with increasing values of the soft SUSY-breaking parameters because the superpartner states begin to decouple. This correlation provides a modest degree of explainability, but one feels that there must exist analytic expressions that can more finely reproduce the dependence of the low-energy observables on the parameters. Indeed if it were possible to infinitely refine such expressions, then the end result would be a set of analytic formulae that would accurately predict all low-energy observables from any given set of input parameters, with no need for time-consuming computation.

For certain observables, $(g-2)_\mu$ for example, some progress can be made following this physics-oriented pathway in which one continually refines expressions with increasing levels of physical complexity. However this is not possible for the majority of observables, for example $\ODM h^2$ which involves multiple scales of physics, complex co-annihilation effects and so forth. For this and indeed most observables the full analytic expressions that would be derived from a purely physical approach would ultimately yield regressors that are quite unintelligible from a physics perspective. Thus, there is clearly a trade-off to be made between the simplicity of analytic expressions and their power as regressors for physical observables. Conversely making expressions easily physically interpretable typically yields very poor regressors, although physical interpretability may be enhanced once accurate analytic regressors have been found. The thrust of this work is that there can be a large improvement in efficiency if one  focusses on the accuracy of analytic regression rather than on its interpretability. However we emphasize that there are further advantages of finding accurate analytic regression formulae beyond mere speed-up, which arises primarily from the fact that such a methodology is still more interpretable than traditional machine learning which produces non-interpretable black-box regressors to reproduce the BSM predictions. For example one can easily use analytic expressions for observables to find interpretable asymptotic formulae in the limit of large or small parameters. In addition perhaps one of the most important and promising aspects of such expressions is that one can take derivatives of them, so they can for example be used in differential programming.

The business of producing analytic expressions that reproduce the output of complicated computations is known as {\it symbolic regression}~\cite{Koza92}. In the physics context, it has most famously been discussed in generality in Ref.~\cite{Udrescu:2019mnk} and for specific applications in Refs.~\cite{Butter:2021rvz,Abel:2022nje,Bartlett:2022kyi,Koksbang:2023sab,Sousa:2023unz,MaurizioEtAl}.
Symbolic regression attempts to provide analytic expressions for the outputs by learning the symbolic formulae that best fit the observed results. It does not attempt to provide any kind of rationale for the expressions it finds (although as in Ref.~\cite{Udrescu:2019mnk} the bank of symbolic expressions which are considered can be motivated by physics), the goal being merely to discover the simplest and most accurate analytic expressions that reproduce the observables in the region of the parameter-space of interest. \resp{As we shall see}\/this is a useful compromise: if we are prepared to forego the physically organised chain of computation that determines the observables at each point in parameter space, then we can \resp{indeed}\/gain much greater analytic power.  

The purpose of this letter is to demonstrate that symbolic regression is a powerful tool for studying BSM physics. As a benchmark model, we will consider the so-called Constrained Minimal Supersymmetric Standard Model (CMSSM), which has a four-dimensional parameter space, consisting of GUT scale degenerate gaugino masses, scalar masses, and universal trilinear coupling, and the electroweak scale Higgs Vacuum Expectation Value (VEV) ratio, denoted respectively as $m_{1/2},~m_0, ~A_0$ and $\tan\beta $. We provide a set of analytical expressions that reproduce the Higgs mass, $m_{H^0}$, the SUSY contribution to the muon anomalous magnetic moment, $\delta(g-2)_\mu$, and the dark matter relic density, $\ODM h^2$, in terms of these parameters (which are available at~\cite{symbolic_regression_bsm_2024} alongside the code that produced them, and the dataset used can be found at~\cite{abdussalam_2024_11366471}). In addition we provide a ``classifier'' $C(m_{1/2},m_0,A_0,\tan\beta)$, which is a function that takes values greater than $0.5$ when a point is physically viable in the sense that it has neutral dark matter, lack of charge and colour breaking minima, and a positive dark matter relic density. As an example application of our methodology, we will demonstrate that by employing these symbolic expressions one may determine the posterior probabilities of the CMSSM extremely rapidly compared to conventional methods.

There are several approaches to symbolic regression that could be considered for this purpose, ranging from evolutionary methods to transformer-based neural networks. A comprehensive overview and comparison of symbolic regression methods is included in Refs.~\cite{defranca,DBLP:journals/corr/abs-1909-05862,DBLP:journals/corr/abs-2006-11287,DBLP:journals/corr/abs-2107-14351,cranmer2023interpretable}. However, the specific properties that we require of a symbolic regressor for this letter, and for BSM more generally, are somewhat specific to BSM physics. Not only is the parameter space often high dimensional, but also it tends to contain fairly focussed regions of interest which are localised around poles and mass-degeneracies, and these need to be captured by the expressions. Consequently, successful training involves a large and relatively fine multidimensional set of training data. This excludes the most commonly used symbolic regression packages (\eg PySR) and favours Operon (and its Pythonically wrapped version, PyOperon)~\cite{Burlacu:2020:GECCOcomp}, due to its highly efficient vectorised structure and low memory footprint. \\

\vspace{-0.2cm} 
\paragraph*{Analytic expressions for the MSSM.---}Operon is a framework for symbolic regression based on Genetic Programming (GP), which is an evolutionary method in which each individual in the population is an {\it expression tree} that represents a symbolic expression built from a bank of pre-chosen functions. Evolution of the population is then simulated by repeated cycles of {\it selection}, {\it breeding}, and {\it mutation}. The selection probability for breeding is governed by a loss-function for each individual, which is determined from the properties of the symbolic formula that is generated by its expression tree. 

Obviously, these properties include closeness to the training data (which in this study is a set of $10^5$ predetermined CMSSM points), but also the properties of the expression itself. The \texttt{SymbolicRegressor} module of PyOperon runs the GP loop with two objectives: one of several possible regression metrics and the ``length'' of the expression itself. The regression metric is a user-defined hyperparameter that can be chosen from one of the following: ``mean square error'', ``mean average error'', ``$r^2$'', and ``normalised mean square error''. The ``length'' is simply the number of characters (string length) of the mathematical expression. During the GP loop, the population is evaluated on these two metrics, with the best individuals being those inhabiting the corresponding Pareto front. At the end of the run (\ie after the specified number of generations/budget is exhausted), \texttt{SymbolicRegressor} returns all the Pareto front individuals.

To optimise the regression metrics, an Optuna loop was implemented in our analysis. This uses a Tree Parzen Estimator (TPE), a Bayesian optimisation algorithm, to determine the best hyperparameters in the algorithm, which, as well as the regression metrics, include for example population size, population initialisation, and so forth (see Ref.~\cite{Burlacu:2020:GECCOcomp}).
As a ``figure of merit'' for a particular choice of hyperparameters, we use the relative error of each of the three observables for each individual in the final Pareto front. Denoting the observables generically as $y$, this is given by 
$
|y_{\rm pred} - y_{\rm true}|/|y_{\rm true}|,
$
where $y_{\rm true} $ are points taken from the validation set (also composed of $10^5$ points). 

The relative error was preferred over other more common regression metrics, such as $r^2$, because the latter tend to be biased toward higher nominal values, which can be problematic for observables that span multiple orders of magnitude. The average of the relative errors was taken, weighted by ``flattening weights'' which level the distribution to ensure that the regressor performs well over the whole range of values of the observable, thus preventing it from focusing on the most common values of the observable. Therefore, during training, because Operon does not use sample weights to produce weighted averages when computing the loss function, at each Optuna iteration, the training data were resampled without replacement according to the ``flattening weights'', which ultimately reduced our useable dataset from $10^5$ to $10^4$ data points, while ensuring a mostly flat distribution of the target observable. 

Such weighting is an important feature especially for $\ODM h^2$ because its values range over many orders of magnitude, and, as we shall see, it poses by far the most challenging regression problem in this study. 
In fact, without resampling the training data to produce a flat distribution the symbolic regressor was unable to accurately map the physical region of interest $\mathcal{O}(\ODM h^2)\sim 0.1$, due to the relative scarcity of points with values in that range. By producing a flatter $\mathcal{O}(\ODM h^2)$ distribution during training, this was greatly improved, but there was still considerable contamination in the range of physical interest. Further significant improvement was achieved by allowing a larger budget and a larger maximal tree-size for both $\ODM h^2$ and $m_{H^0}$, and rerunning the Optuna loop to produce their final expressions. Meanwhile, for $\delta(g-2)_\mu$ and the classifier it was sufficient to keep the smaller expressions obtained with a smaller budget.

\begin{figure*}[t]
  \centering
  \vspace{-1cm} 
  \includegraphics[scale=0.38]{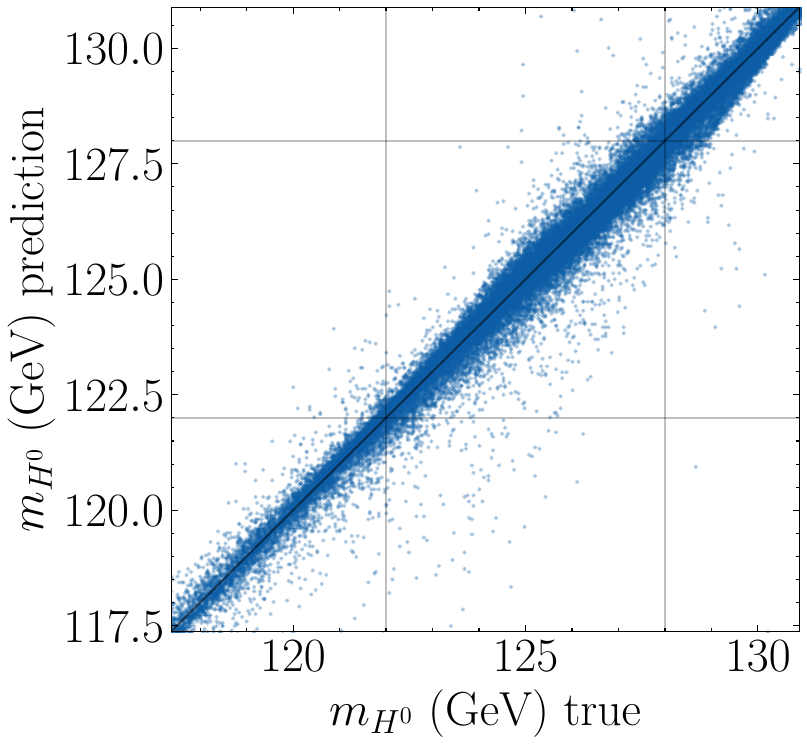}
  ~~~~~~\includegraphics[scale=0.38]{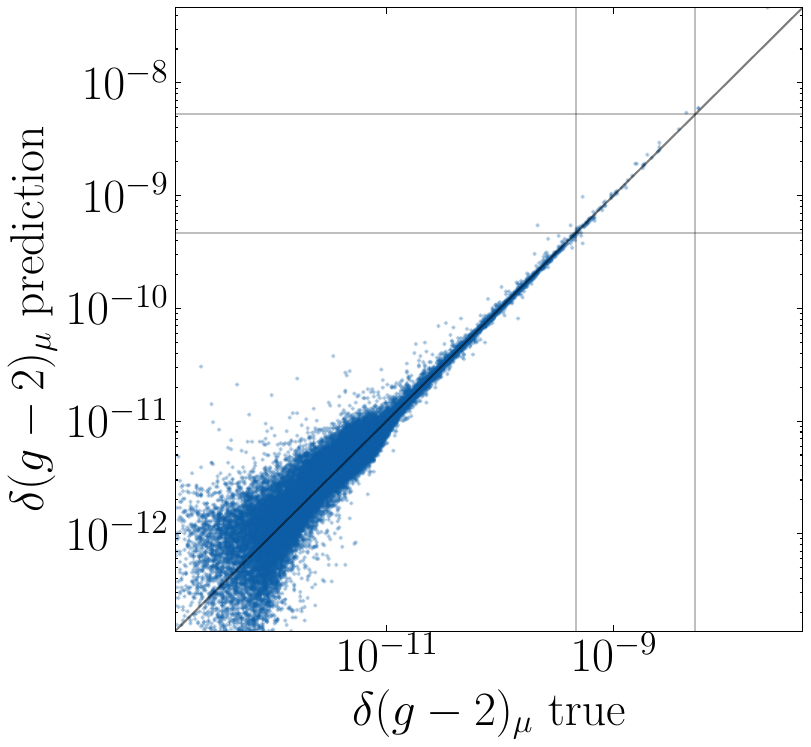}
  ~~~~~~\includegraphics[scale=0.38]{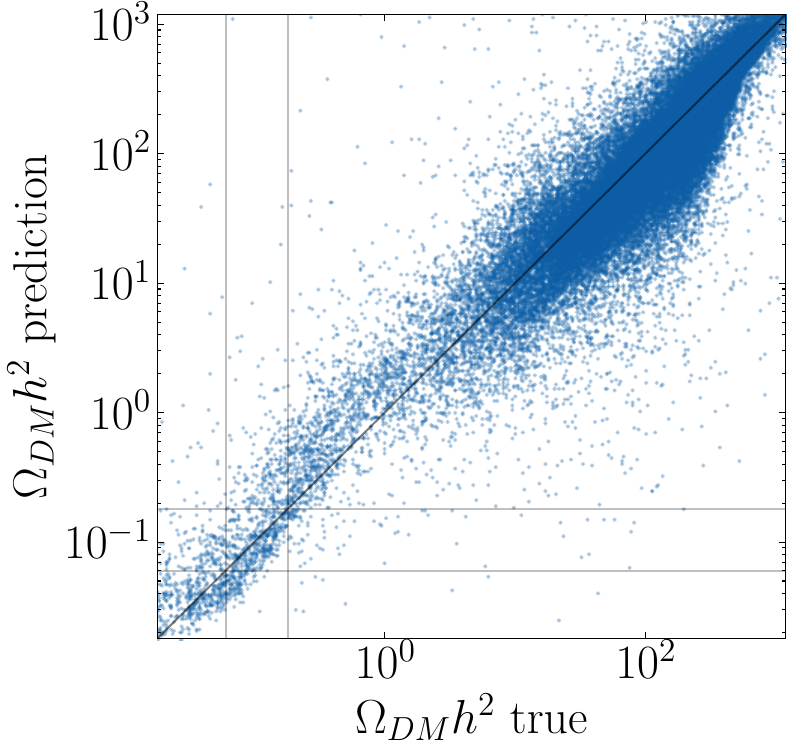}
  \includegraphics[scale=0.38]{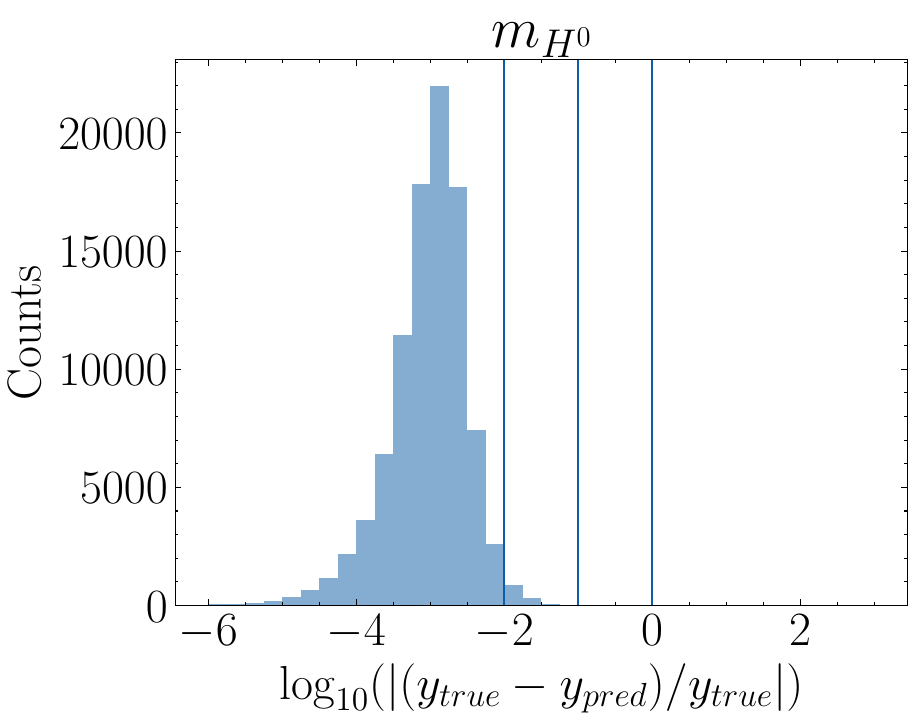}
  \includegraphics[scale=0.38]{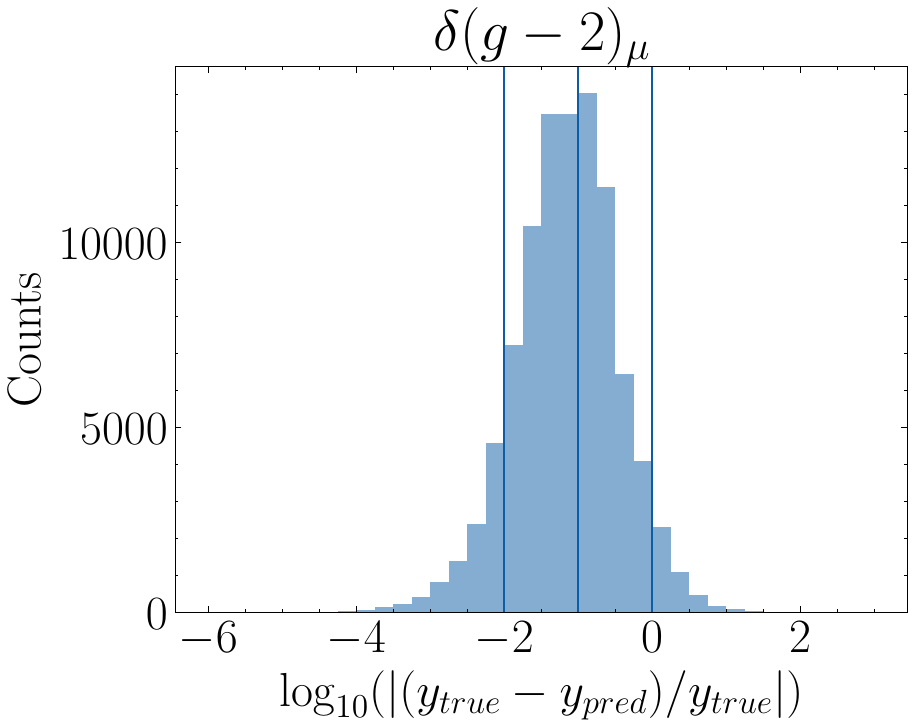}
  \includegraphics[scale=0.38]{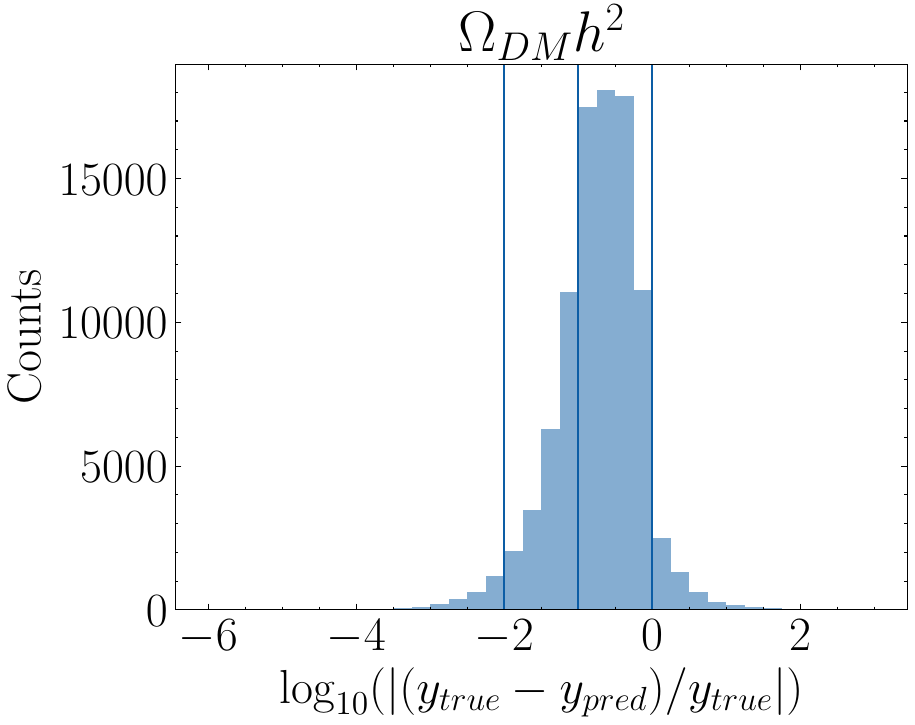}
  \caption{Symbolic regressors' performance on the test set. Upper panels: ``true vs. prediction'' scatter plots, the solid lines marking the boundary of the physically viable values of the observables. Lower panels: distribution of relative errors, the vertical lines offering a visual guide for relative errors of order 1\%, 10\%, and 100\%.}
  \label{fig:true-pred-and-rel-errors}
\end{figure*}

To evaluate the quality of the symbolic regressors that were obtained with this procedure, we present in \cref{fig:true-pred-and-rel-errors} the ``true vs. prediction'' scatter plots (upper panels) and the distribution of the relative errors of the predictions (lower panels), produced using the test set. It can be seen that for the Higgs mass the regression is very accurate throughout all the values, with virtually every point having a relative error below 1\%. For $\delta (g-2)_\mu$ the relative errors are greater, but importantly remain diminishingly small in the region of physical viability (where $\delta(g-2)_\mu$ is enhanced). (The distribution of points is typical of a regressor with a constant absolute error.) For $\ODM h^2$ relative errors are on average above 10\%. This is actually better than the 20\% {\it theoretical} error that is typically assigned to $\ODM h^2$ in global fit analyses. Despite this, the symbolic regressor is capable of producing viable estimates in the physical region of interest. It is important to note that each scatter plot shows $10^5$ points of the test set, and the points with poor predictions in the upper panels of \cref{fig:true-pred-and-rel-errors} are actually very few, constituting only a small scattered minority of the test set (as is evident in the lower panels of \cref{fig:true-pred-and-rel-errors}).

\begin{figure*}[t]
  \centering
  \includegraphics[scale=0.42]{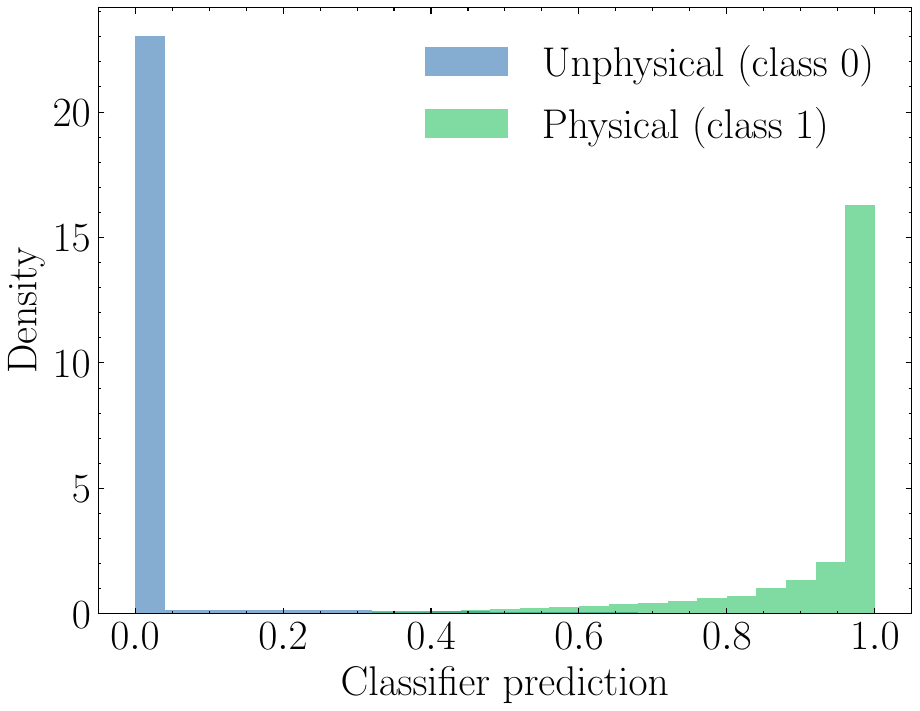}
  ~~~~~~\includegraphics[scale=0.42]{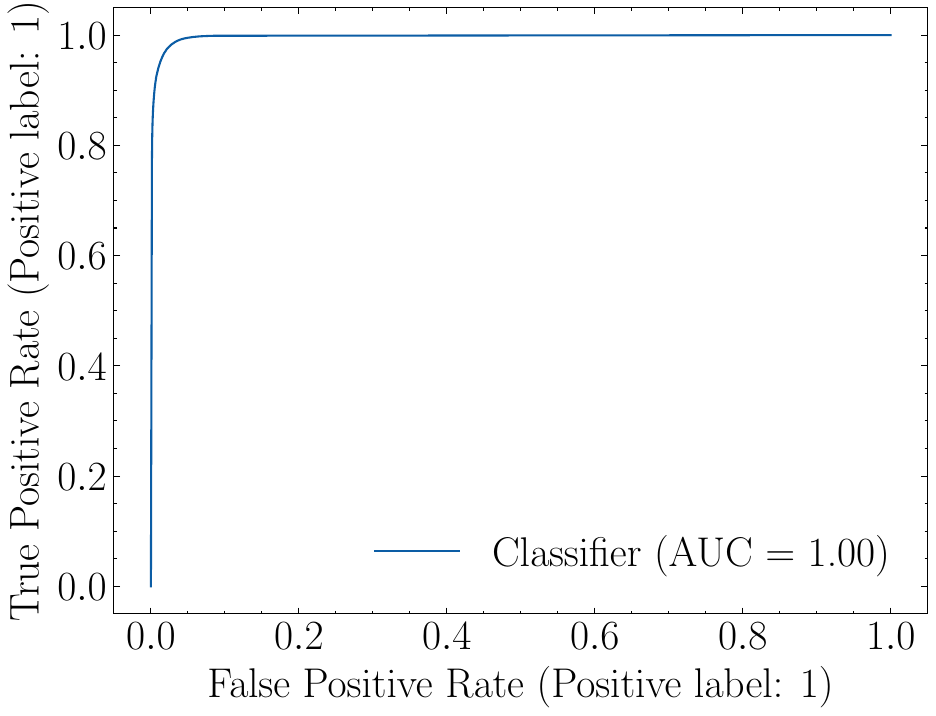}
  \caption{\resp{Performance of the classifier symbolic regressor on the test set. Left panel shows the output of the classifier where class 0 (in blue) are physically disallowed points and class 1 (in green) are allowed points. The right panel shows the ROC curve which makes clear the excellent performance of the classifier.  } }
  \label{fig:classifier}
\end{figure*}

\resp{Regarding the classifier regressor, which can be used to discriminate unphysical from physical points in order to allow rapid rejection of the former, the performance is shown in the two panels of Fig.~\ref{fig:classifier}. Clearly the performance of this classifier regressor is extremely good! }

\paragraph*{Application: global fits.---}
Generally, to make
convergent and statistically robust global fits, say using state-of-the-art nested sampling
algorithms, the main difficulty is the need to draw samples at each iteration of the algorithm, 
until a point is found that has a likelihood greater than that of the lowest point within the likelihood-sorted live points. There 
are various approaches to achieving this 
with varying costs to compute the usually simulation-based or scientific software
package-based 
likelihoods. Hence, we would now like to demonstrate that a complete or partial replacement with symbolic expressions of the high energy physics packages employed for making BSM global analyses is a solution to the resource problems that are associated with the stringent experimental limits and the non-observation of new fundamental particles. 

For our benchmark model, the CMSSM, the observables  $m_{H^0}$, $\ODM h^2$, and $\delta (g-2)_\mu$ can be computed for each point in the parameter space via particle spectrum generators such as SPheno~\cite{Porod:2003um} and particle dark matter packages such as MicrOMEGAs~\cite{Belanger:2001fz}. We use the most precise Higgs boson measurement, $m_{H^0} = 125.04 \pm 0.12$~GeV~\cite{CMS-PAS-HIG-21-019} and $\ODM h^2 = 0.12 \pm 0.0012$~\cite{Planck:2018vyg} but include the aforementioned theoretical uncertainty of 20\% in predicting the relic density. Thus, $\ODM h^2 = 0.12 \pm 0.02$ is used for the fits. The discrepancy between the high precision Standard Model (SM) prediction for the muon anomalous magnetic moment~\cite{Aoyama:2020ynm} and the experimental measurements~\cite{Muong-2:2023cdq} which we adopt for the fits is $\delta (g-2)_\mu = (249 \pm 48) \times 10^{-11}$. These constitute the set of data, $\underline{d} = \{ \mu_i \pm \sigma_i \}$  that we use to make the global fit, where $\mu_i$ and $\sigma_i$ represent the measurement central values and uncertainties for the above observables (with $i=1,2,3$).

The CMSSM parameters were sampled from uniformly distributed prior probability densities, which were respectively within $[0, 10]$~TeV for $m_{1/2}$ and $m_0$, $[-6, 6]$~TeV for $A_0$, and $[1.5, 50]$ for $\tan\beta$. For each point in the parameter space, $\underline{\theta} = \{ m_{1/2}, m_0, A_0, \tan \beta\}$, the likelihood is estimated as 
\beq 
p(\underline{d} | \underline{\theta}) ~=~ \prod_{j=1}^3 \, \frac{1}{\sigma_j \sqrt{2\pi}} \, exp\bigg\{ \frac{-(y_{\rm pred}^j - \mu_j)^2}{2\sigma_j^2} \bigg\}~,
\eeq 
where again $y_{\rm pred}^j$ represents the predictions for our three observables. \resptwo{We study the results for two distinct implementations of nested sampling by MultiNest and Dynesty.} We used version 3.10 of MultiNest with 4000 live points in the nested importance sampling mode and with tuning parameters chosen as $\mbox{\tt mmodal} = 1$, $\mbox{\tt ceff} = 0$, $\mbox{\tt efr} = 1.8$, $\mbox{\tt tol} = 0.5,$ and $\mbox{\tt seed} = -1.$ \resptwo{For Dynesty, we used version 2.1.4, and ran the \texttt{DynamicNestedSampler} with the default 500 live points, no \texttt{bootstrap}, with \texttt{pfrac=1} for posterior estimation.}

\begin{figure*}
  \centering
\includegraphics[width=0.8\textwidth]{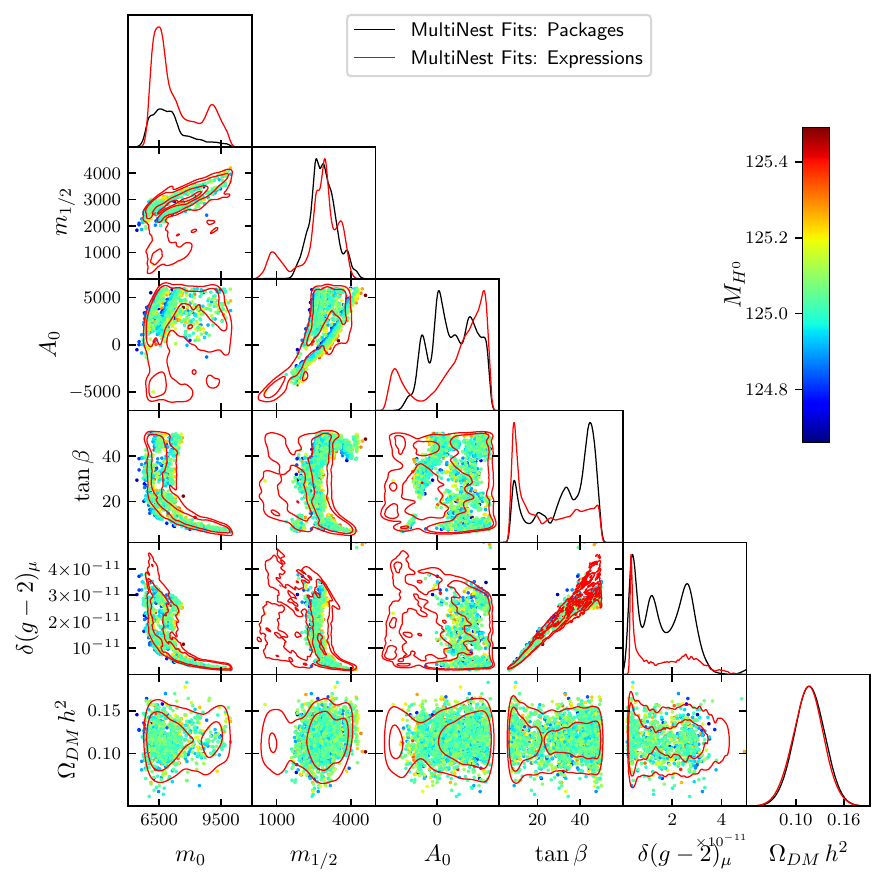}
  \caption{The posterior distributions obtained by MultiNest for global fits of the CMSSM parameters to $m_{H^0}$, $\delta (g-2)_\mu$, and $\ODM h^2$, using solely package-based (2-dimensional scatter plots on the off-diagonal entries, and black lines on the 1-dimensional plots) versus solely expression-based (the red lines) approaches. Mass dimensionful parameters are in GeV.}
  \label{fig:tri_mh}
\end{figure*}

\begin{figure*}
  \centering
\includegraphics[width=0.8\textwidth]{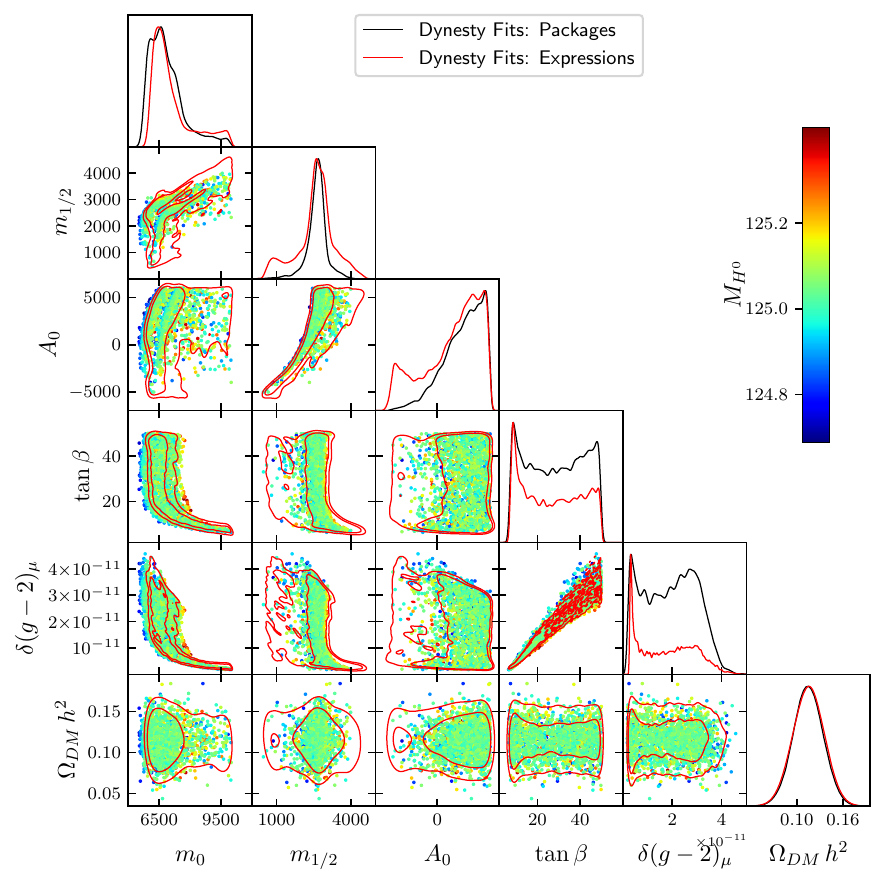}
  \caption{\resptwo{The posterior distributions obtained by Dynesty for global fits of the CMSSM parameters to $m_{H^0}$, $\delta (g-2)_\mu$, and $\ODM h^2$, using solely package-based (2-dimensional scatter plots on the off-diagonal entries, and black lines on the 1-dimensional plots) versus solely expression-based (the red lines) approaches. Mass dimensionful parameters are in GeV.}}
  \label{fig:tri_mh_dyn}
\end{figure*}

We find that the global fit analysis using symbolic expressions compares exceedingly well with that made using the conventional method (packages-based). \resptwo{To illustrate,  Figs.~\ref{fig:tri_mh},~\ref{fig:tri_mh_dyn} show the posterior distributions of the CMSSM parameters fit to our three observables using MultiNest and Dynesty respectively, plotted using GetDist~\cite{Lewis:2019xzd}. The 65\% and 95\% Bayesian probability contour lines (in red on the 2-dimensional plots, and labelled ``Expressions'' in the legend) represent the 2-dimensional posterior distributions from the symbolic regression fit, while the 2-dimensional scatter plots (with points coloured according to the Higgs mass, and labelled ``Packages'' on the legends) represent conventional fits in which all the observables are computed with packages. These results are clearly in excellent agreement, confirming that the dynamical features of the symbolic expressions correctly approximate those of the packages.}

\resptwo{We note that the MultiNest fit using packages returned some evidence for a second modality for $m_0\sim 1$ TeV, with slightly higher values of $\delta(g-2)_\mu$. This region has been omitted in the plots above for ease of readability, as Dynesty with packages also missed it. While the goal of our paper is not to provide a realistic global fit to the CMSSM, we also believe that this region would not survive current collider limits, and therefore does not change the discussion presented herein. We believe that the expressions did not capture this mode because the data-set (which was produced by random sampling) used to produce the expressions had very few phenomenologically viable points with high likelihood in that region of the parameter space.}

\paragraph*{Discussion.---} 

\resptwo{Given these results , we can conclude that symbolic regression can be a very powerful tool where analytic results are absent or hard to employ. While it may be possible to analytically approximate the form of the Higgs mass for example, observables such as $\ODM h^2$ have no known analytic formula, and indeed it is difficult to see how one could ever be derived. 
Symbolic regression can successfully provide predictive expressions for such observables.}

Generally, one anticipates a huge reduction in the computational resources required to perform a global fit using symbolic regression: for example, we find that \resptwo{two orders of magnitude less CPU time are required} for the CMSSM global fit when all three of our observables are computed using the symbolic expressions. In this comparison we used simple CPU-time and did not factor in the time required to generate the data set nor the time required to train PyOperon. However note that our time comparison is between two methods for making fits, and the training times constitute an initial investment to generate the symbolic expressions which can then be re-used for future studies.

It is interesting to note that symbolically regressed expressions can be in conflict, and some care must be taken in using them according to the physical problem. \resptwo{When this is the case, we argue that it reflects the tension between constraints in the underlying BSM candidate.} In fact, this partly motivated our choice of these three observables for this particular study. Indeed, studying our training data points in a frequentist fashion reveals that good $\ODM h^2$ values are mostly obtained for Higgsino dark matter, whereas good $\delta(g-2)_\mu$ values are mostly obtained for Bino dark matter. In other words, good values for these two observables are relatively rare in the CMSSM parameter space, and the most populated regions do not overlap. 
Thus our symbolic expressions for $\Omega h^2$ are not sensitive to those very rare points that {\it also} have the required $\delta(g-2)_\mu$ contribution, but are instead swamped by Higgsino dark matter points. \resptwo{The existence of alternative modalities is difficult for the expressions (and also Dynesty) to pick up simply because the original data-set had very few points with high likelihood. Therefore if one wished to refine the global fit using symbolic expressions one should proceed by first including into the {\it classifier regressor} the requirement that the dark matter candidate should be Bino-like. In order to do this one would use the methodologies developed in \cite{Romao:2024gjx} in order to enrich the data set with points of phenomenological interest. This and similar refinements will be the subject of future work. By contrast, it is already remarkable that such accurate expressions could be obtain from a non-ideal data-set (produced, recall, by random sampling) as can be confirmed by the posteriors presented above.  }

We should briefly remark on how the method scales with complexity, in particular with an increase in the number of parameters and constraints. First we note that as we saw in Fig.~\ref{fig:classifier} the symbolic classifier is extremely effective for the constraints that are being applied. This includes charge and colour breaking minima, charged dark matter and negative $\ODM h^2$. However in the parameter space note that these are effectively many constraints as there are many flat directions in parameter space that can yield CCB minima. Therefore given this and the remarkable strength of the classifier it is unlikely that additional constraints would lead to a bottle-neck. On the other hand the fitting of more observables may become more difficult if the associated regressors have poor quality, for example if the observables in question behave more like $\ODM h^2$ (which is hard to regress) rather than the Higgs mass (which is easy). Several observables with poor regressors would in turn impact the quality of the fits. This is indeed one of the important reasons for seeking good regressors rather than interpretable ones. Of course for other studies the methodology can be adapted accordingly.

Regarding the applicability of these methods, we see them as exploratory tools that can be used in conjunction with more conventional methods in order to map out BSM parameter spaces. To appreciate the potential impact of this method, note that there are multiple physics cases where calculating the observables for a {\it single point} can take minutes -- as an example reference Ref.~\cite{Diaz:2024yfu} quotes 120 seconds to evaluate a single point by conventional methods. If a good regressor can be found for these observables then it can be used as a replacement in subsequent analyses which are focussed on different physical aspects, without the need to recompute the observables for each new point in the study, or to attempt to exhaustively scan a parameter space (which is of course impossible if each point takes 120 seconds). In this sense the symbolic regression approach is akin to ``amortised inference'' which seeks to re-use results of a previous analysis by training a generative model that can be used for subsequent studies.

As a final comment, we remark that as well as
offering a different and very efficient way of analysing BSM models, symbolic regression 
 opens up several new avenues for analysing high energy physics models. As an example, 
 it is interesting to note that it may now be possible to perform global fit analyses on a quantum computer in a manner analogous to that in Ref.~\cite{Criado:2022aoo}, but instead by directly encoding the symbolic expressions for the observables in the quantum circuit. This kind of analysis would be out of the question using the conventional calculational route due to the impossibility of fully encoding the required chain of computation in a quantum circuit. Another promising application of our approach is to further the potential of differentiable and probabilistic programming in BSM studies, where the symbolic expressions can replace the black-box imposed by the computational packages.\\

\vspace{-0.2cm} 
\noindent 
We are extremely grateful to Bogdan Burlecu and Gabriel Kronberger for extensive guidance with Operon. We would like to thank Miles Cranmer and Pedro Ferreira for help and discussions. SA$_{1}$ and SA$_{2}$ thank CERN-TH and SA$_{1}$ thanks the Institute for Theoretical Physics at Heidelberg University for hospitality extended during the initial stages of this work. SA$_2$ and MCR are supported by the STFC under Grant No. ST/T001011/1. This work was performed using resources provided by the Cambridge CSD3, provided by Dell EMC and Intel using Tier-2 funding from EPSRC grant EP/T022159/1, and DiRAC funding from the STFC.

\vspace{-0.4cm}

\bibliography{paper} 

\end{document}